\documentclass[reprint,aps,prd,twocolumn,notitlepage,showpacs,nofootinbib,preprintnumbers,superscriptaddress]{revtex4-1}
\usepackage{amsmath}
\usepackage{amsfonts}
\usepackage{amssymb}
\usepackage{latexsym}
\usepackage{enumerate}
\usepackage{color,xcolor}
\usepackage{graphicx}
\usepackage{multirow}
\usepackage{bm}
\usepackage{epsfig}
\usepackage[english]{babel}
\usepackage{hyperref}
\usepackage{times}
\usepackage{comment}
\usepackage{epstopdf}

\def\d{\mathrm{d}}
\def\dom{\left(\d\theta^2+\sin^2\theta \d\phi^2\right)}
\def\rh{r_\mathrm{H}}
\def\max{\mathrm{max}}
\def\min{\mathrm{min}}
\def\En{\mathcal{E}}
\def\vx{\mathbf{x}}
\def\vv{\mathbf{v}}

\begin{document}
%\preprint{\vbox{ \hbox{}   \hbox{} }}

\title{Dark matter distributions around Schwarzschild-like black holes in bumblebee and Kalb-Ramond models}
\author{Ming-Hong Yu}
\author{Towe Wang}
\email[Electronic address: ]{twang@phy.ecnu.edu.cn}
\affiliation{School of Physics and Electronic Science, East China Normal University, Shanghai 200241, China\\}
\date{\today\\ \vspace{1cm}}
\begin{abstract}
A central black hole can attract a dark matter cluster and generate a spike in the density profile, as demonstrated by detailed analysis of Schwarzschild and Kerr black holes in the past. Do different black holes attract dark matter differently? To get a fair answer to this question, we customize a relativistic framework to grow general static spherical black holes in dark matter halos and investigate how deviations from the Schwarzschild geometry modify the dark matter spike for the first time. The framework is applied to a class of Schwarzschild-like black hole solutions in Lorentz-violated gravity models --- one in the bumblebee model and two in the Kalb-Ramond model. For these black holes, the answer is no if initially the dark matter has a constant distribution, but the answer is yes if it has a Hernquist profile initially.
\end{abstract}

%\pacs{}

\maketitle

%\tighten

%%%%%%%%%%%%%%%%%%%%%%%%%%%%%%%%%%%%%%%%%%

%\tableofcontents

\section{Introduction}\label{sect-intro}
The dark matter, one of the main ingredients in the Universe, has been investigated for almost a century but remains an enigma \cite{Bertone:2004pz}. The details of its distributions in galaxies are crucial for both astronomical detections and model construction \cite{Hernquist:1990be,Zhao:1995cp}. As demonstrated by both nonrelativistic and relativistic analyses \cite{Gondolo:1999ef,Sadeghian:2013laa,Ferrer:2017xwm}, when a black hole appears at the center of a dark matter halo, a spike will emerge in the density profile.\footnote{However, different results were reported in Refs. \cite{Ullio:2001fb,Merritt:2003qk,Bertone:2024wbn}.} So far as we know, concrete analysis has been performed only for Schwarzschild and Kerr black holes. For other black holes, the spike is expected to exist and modeled by analytic functions \cite{Capozziello:2023rfv,Shen:2023erj,Zhang:2025mdl}, but the details are largely omitted.

In the present paper, we adapt the relativistic adiabatic growth framework established in Ref. \cite{Sadeghian:2013laa} to general static spherical black holes, and then we implement it to study the distribution of dark matter around three Schwarzschild-like black holes in detail. The first black hole is a solution in a bumblebee gravity model \cite{Casana:2017jkc}. The other two black holes are solutions in a Kalb-Ramond model \cite{Yang:2023wtu,Liu:2024oas}. As we will demonstrate in this paper, if one cares only about the gravitational fields, these solutions are equivalent. They have a geometry akin to the Schwarzschild black hole, but the space is contracted or stretched in the radial direction by a deformation parameter. The deformation parameter is related to the spontaneous breaking of Lorentz symmetry in bumblebee and Kalb-Ramond models.

As concrete examples, we calculate the final density profile of dark matter around the Schwarzschild-like black holes, assuming that the dark matter has either a constant distribution or a Hernquist density profile initially. When the initial distribution of the dark matter is the constant distribution, we find Schwarzschild and Schwarzschild-like black holes of the same radius attract dark matter identically and yield exactly the same density profile. When the initial density profile of the dark matter is the Hernquist profile, we find the black holes attract dark matter differently and the final dark matter density is dependent on the value of the deformation parameter. Particularly, depending on the sign of the deformation parameter, the dark matter spike is sharper or blunter near the Schwarzschild-like black holes than it is near the Schwarzschild black hole.

The layout of this paper is as follows. In Sec. \ref{sect-sph}, we set up the framework for studying dark matter distribution around a general static spherical black hole, covering the conserved quantities of dark matter particles in Sec. \ref{subsect-cons}, the mass-current density four-vector of dark matter in Sec. \ref{subsect-mcur}, the adiabatic invariance of dark matter distribution in Sec. \ref{subsect-adinvar}, and the mass density profile of dark matter in Sec. \ref{subsect-dens}. Then this framework is applied to Schwarzschild-like black holes in bumblebee and Kalb-Ramond models in Sec. \ref{sect-Slike}. After recalling three such black hole solutions and verifying their equivalence, we simulate the density profile of dark matter around Schwarzschild-like black holes starting from a constant distribution in Sec. \ref{subsect-con} and from a Hernquist distribution in Sec. \ref{subsect-Hern}. Finally, in Sec. \ref{sect-sum}, we summarize our main results. In Appendix \ref{app-nzell}, we present a linear approximation of the difference between dark matter densities around Schwarzschild-like and Schwarzschild black holes when the deformation parameter is close to zero.

Throughout this paper, we will assume that the dark matter is composed of particles of the same mass $\mu$, its steady-state distribution is of spherical symmetry, the spherical black hole grows adiabatically in the center, and the particles are moving in bound orbits outside the black hole. We will adopt the metric signature $(-+++)$ and express all quantities in natural units $(\hbar=c=1)$.

\section{Growing static spherical black holes in a dark matter cluster}\label{sect-sph}
\subsection{Conserved quantities}\label{subsect-cons}
Let us begin by writing the line element of a spherically symmetric static black hole in the general form
\begin{equation}\label{metric-sph}
\d s^2=-A(r)\d t^2+B(r)\d r^2+R(r)^2\dom,
\end{equation}
where $A(r)$, $B(r)$ and $R(r)$ are determined by specific black hole solutions, but in this section we will leave them as unspecified functions of the radial coordinate $r$. We deliberately choose $r$ rather than $R$ as the radial coordinate so that our framework will have a wider range of applications, especially when $R(r)$ is an elementary function but its inverse is nonelementary.

%Adapted to this line element, the present section will be dedicated to a brief review of the general relativistic analysis in Ref. \cite{Sadeghian:2013laa}.

For a dark matter particle of rest mass $\mu$ moving along timelike geodesics in this spacetime, the Hamiltonian takes the form
\begin{eqnarray}\label{Hamilton}
\nonumber\mathcal{H}&=&\frac{1}{2}g^{\alpha\beta}p_\alpha p_\beta\\
&=&\frac{1}{2}\left[-\frac{p_t^2}{A(r)}+\frac{p_r^2}{B(r)}+\frac{p_{\theta}^2}{R(r)^2}+\frac{p_{\phi}^2}{R(r)^2\sin^2\theta}\right].
\end{eqnarray}
By substitution of $p_\alpha=\partial S/\partial x^\alpha$ into the Hamiltonian, one can write the Hamilton-Jacobi equation $\partial S/\partial\lambda+\mathcal{H}=0$ concretely as
\begin{eqnarray}\label{HJE}
\nonumber-\frac{\partial S}{\partial\lambda}&=&\frac{1}{2}g^{\alpha\beta}\frac{\partial S}{\partial x^\alpha}\frac{\partial S}{\partial x^\beta}\\
\nonumber&=&\frac{1}{2}\left[-\frac{1}{A(r)}\left(\frac{\partial S}{\partial t}\right)^2+\frac{1}{B(r)}\left(\frac{\partial S}{\partial r}\right)^2\right.\\
&&\left.+\frac{1}{R(r)^2}\left(\frac{\partial S}{\partial\theta}\right)^2+\frac{1}{R(r)^2\sin^2\theta}\left(\frac{\partial S}{\partial\phi}\right)^2\right],
\end{eqnarray}
in which $S$ is Hamilton's principal function and $\lambda$ is the affine parameter. The Hamiltonian \eqref{Hamilton} is independent of $\lambda$ and subject to the constraint $\mathcal{H}=-\mu^2/2$. Moreover, the Hamiltonian does not depend explicitly on coordinates $t$ and $\phi$, so their conjugate momenta are conserved. As a result, we immediately find three conserved quantities: the energy per unit mass $\En$, the $z$-component of the angular momentum per unit mass $L_z$, and the rest mass of the dark matter particle $\mu$. In terms of the contravariant four-momentum $p^\alpha$ and in accordance with Eq. \eqref{metric-sph}, they can be expressed as
\begin{eqnarray}
\label{cons-E-Lz}\En&=&-u_t=\frac{1}{\mu}A(r)p^t,\quad L_z=u_\phi=\frac{1}{\mu}R(r)^2\sin^2\theta p^\phi,\\
\label{cons-mu}\nonumber\mu^2&=&-g_{\alpha\beta}p^\alpha p^\beta=A(r)(p^t)^2-B(r)(p^r)^2-R(r)^2(p^{\theta})^2\\
&&\qquad\qquad\qquad-R(r)^2\sin^2\theta(p^{\phi})^2.
\end{eqnarray}
Note that the four-velocity $u^\alpha=\d x^\alpha/\d\lambda$ is related to the four-momentum by $u^\alpha=p^\alpha/\mu$.

The conservation of $p_t$, $p_\phi$ and $\mathcal{H}$ leads to several partial differential equations of $S(t,r,\theta,\phi,\lambda)$. They are
\begin{equation}\label{pdS}
\frac{\partial S}{\partial t}=-\mu\En,\quad\frac{\partial S}{\partial\phi}=\mu L_z,\quad\frac{\partial S}{\partial\lambda}=\frac{1}{2}\mu^2.
\end{equation}
Inserting them into Eq. \eqref{HJE}, we obtain
%\begin{equation*}
%\frac{1}{2}\mu^2+\frac{1}{2}\left[-\frac{\mu^2\En^2}{A(r)}+\frac{1}{B(r)}\left(\frac{\partial S}{\partial r}\right)^2+\frac{1}{R(r)^2}\left(\frac{\partial S}{\partial\theta}\right)^2+\frac{\mu^2L_z^2}{R(r)^2\sin^2\theta}\right]=0,
%\end{equation*}
\begin{eqnarray}\label{separation}
\nonumber&&\frac{1}{\mu^2}\left(\frac{\partial S}{\partial\theta}\right)^2+\frac{L_z^2}{\sin^2\theta}\\
&=&R(r)^2\left[-1+\frac{\En^2}{A(r)}-\frac{1}{\mu^2B(r)}\left(\frac{\partial S}{\partial r}\right)^2\right].
\end{eqnarray}
%The solution of Eq.\eqref{HJE} have the form as following
%\begin{equation}\label{HJEsol}
%	S=-\mathcal{H}\lambda-\En t+L_z\phi+S_r+S_\theta.
%\end{equation}
This equation has a solution if and only if both sides are equal to a nonnegative constant, which will be denoted by $L^2$. In other words,
\begin{eqnarray}
\label{left}\frac{1}{\mu^2}\left(\frac{\partial S}{\partial\theta}\right)^2+\frac{L_z^2}{\sin^2\theta}&=&L^2,\\
\label{right}R(r)^2\left[-1+\frac{\En^2}{A(r)}-\frac{1}{\mu^2B(r)}\left(\frac{\partial S}{\partial r}\right)^2\right]&=&L^2.
\end{eqnarray}
It will be useful to rewrite them as
\begin{eqnarray}
\label{utheta}u^\theta&=&\pm\frac{1}{R(r)^2}\left(L^2-\frac{L_z^2}{\sin^2\theta}\right)^{1/2},\\ \label{ur}u_r&=&\pm\sqrt{\frac{B(r)}{A(r)}}V(r)^{1/2},\\
\label{V}V(r)&=&\En^2-A(r)\left[1+\frac{L^2}{R(r)^2}\right].
\end{eqnarray}
Please keep Eq. \eqref{V} in mind because it will be frequently used in this section. At the same time, Eq. \eqref{left} can be reexpressed in terms of the four-momentum as
\begin{equation}\label{cons-L}
L^2=\frac{1}{\mu^2}R(r)^4\left[(p^\theta)^2+\sin^2\theta(p^\phi)^2\right]
\end{equation}
or in terms of the four-velocity as
\begin{equation}\label{cons-L}
L=R(r)\left[R(r)^2(u^\theta)^2+R(r)^2\sin^2\theta(u^\phi)^2\right]^{1/2}.
\end{equation}
Thereby we find the fourth conserved quantity: the total angular momentum per unit mass $L$.

Some remarks are in order here. In this subsection, we interpret the conserved quantity $\En$ as energy per unit mass, and the conserved quantities $L_z$ and $L$ as angular momentum per unit mass. One should be cautious that the interpretation is correct only if $A(r)\rightarrow1$, $B(r)\rightarrow1$, $R(r)\rightarrow r$ at spatial infinity $r\rightarrow\infty$. Otherwise more care should be taken, though $\En$, $L_z$ and $L$ are still conserved quantities.

\subsection{Mass-current density four-vector}\label{subsect-mcur}
The mass-current density four-vector of the dark matter is defined as $J^\alpha=\rho\bar{u}^\alpha$, where $\bar{u}^\alpha$ is the local average four-velocity, and $\rho$ is the mass density measured in the local rest frame. Denoting the relativistic distribution function of the dark matter particles as $f^{(4)}(x,p)$ in the eight-dimensional phase space $(x^\alpha,p^\alpha)$, one can express the mass-current density four-vector in the integral form \cite{Sadeghian:2013laa}
\begin{equation}\label{mcur}
J^\alpha(x)=\int f^{(4)}(x,p)\frac{p^\alpha}{\mu}\sqrt{-g}\d^4p.
\end{equation}
This integration is performed over four-momentum $(p^t,p^r,p^\theta,p^\phi)$, which is related to $(\En,L^2,L_z,\mu)$ by Eqs. \eqref{cons-E-Lz}, \eqref{cons-mu} and \eqref{cons-L}. As elaborated in the previous subsection, $\En$, $L^2$, $L_z$ and $\mu$ are conserved quantities. Therefore, the integration region is more regular after the transformation from $(p^t,p^r,p^\theta,p^\phi)$ to $(\En,L^2,L_z,\mu)$, which is a four-to-one mapping according to Eqs. \eqref{utheta} and \eqref{ur}. Including a factor of $4$, it converts the four-momentum volume element to
\begin{equation}
\sqrt{-g}\d^4 p=4\sqrt{-g}|\mathcal{J}|^{-1}\d\En\d L^2 \d L_z\d\mu,
\end{equation}
in which $\sqrt{-g}=\sqrt{A(r)B(r)}R(r)^2\sin\theta$, and the Jacobian
\begin{eqnarray}
\nonumber\mathcal{J}&=&\frac{\partial(\En,L^2,L_z,\mu)}{\partial(p^t,p^r,p^\theta,p^\phi)}\\
\nonumber&=&
%	\left|\begin{array}{cccc}
%		-\frac{g_{tt}}{\mu}-\frac{\En}{\mu}\frac{\partial\mu}{\partial p^t}\ &-\frac{\En}{\mu}\frac{\partial\mu}{\partial p^r}\ &-\frac{\En}{\mu}\frac{\partial\mu}{\partial p^\theta}\ &-\frac{\En}{\mu}\frac{\partial\mu}{\partial p^\phi}\\[5pt]
%		-\frac{2L^2}{\mu}\frac{\partial\mu}{\partial p^t}\ &-\frac{2L^2}{\mu}\frac{\partial\mu}{\partial p^r}\ &\frac{2p^\theta(g_{\theta\theta})^2}{\mu^2}-\frac{2L^2}{\mu}\frac{\partial\mu}{\partial p^\theta}\ &\frac{2p^\phi(g_{\theta\theta})^2\sin^2\theta}{\mu^2} -\frac{2L^2}{\mu}\frac{\partial\mu}{\partial p^\phi}\\[5pt]
%		-\frac{L_z}{\mu}\frac{\partial\mu}{\partial p^t}\ &-\frac{L_z}{\mu}\frac{\partial\mu}{\partial p^r}\ &-\frac{L_z}{\mu}\frac{\partial\mu}{\partial p^\theta}\ &\frac{g_{\phi\phi}}{\mu}-\frac{L_z}{\mu}\frac{\partial\mu}{\partial p^\phi}\\[5pt]
%		\frac{\partial\mu}{\partial p^t}\ &\frac{\partial\mu}{\partial p^r}\ &\frac{\partial\mu}{\partial p^\theta}\ &\frac{\partial\mu}{\partial p^\phi}
%	\end{array}\right|
%	\\[6pt]
%	&=&\frac{1}{\mu^3}
%	\left|\begin{array}{cccc}
%		-g_{tt}\quad&0\quad&0\quad&0\\[5pt]
%		0\quad&0\quad&2u^\theta(g_{\theta\theta})^2\quad&2u^\phi(g_{\theta\theta})^2\sin^2\theta\\[5pt]
%		0\quad&0\quad&0\quad&g_{\phi\phi}\\[5pt]
%		\frac{\partial\mu}{\partial p^t}\quad&\frac{\partial\mu}{\partial p^r}\quad&\frac{\partial\mu}{\partial p^\theta}\quad&\frac{\partial\mu}{\partial p^\phi}
%	\end{array}\right|
%	\\[6pt]
\frac{1}{\mu^3}\left|\begin{array}{cccc}
A(r)\quad&0\quad&0\quad&0\\[5pt]
0\quad&0\quad&2u^\theta R(r)^4\quad&2u^\phi R(r)^4\sin^2\theta\\[5pt]
0\quad&0\quad&0\quad&R(r)^2\sin^2\theta\\[5pt]
-u_t\quad&-u_r&-u_\theta&-u_\phi
\end{array}\right|\\[2pt]
&=&-2\mu^{-3}A(r)R(r)^6u_ru^\theta\sin^2\theta.
\end{eqnarray}
Assembling them together, we find the four-momentum volume element is converted to
\begin{equation}\label{mvol}
\sqrt{-g}\d^4 p=\frac{2\mu'^3}{R(r)^4|u_r||u^\theta|\sin\theta}\sqrt{\frac{B(r)}{A(r)}}\ \d\En\d L^2\d L_z\d\mu'.
\end{equation}

By assumption all the dark matter particles have the same rest mass $\mu$, so $f^{(4)}(x,p)$ is nonzero only on the mass shell $g_{\alpha\beta}p^\alpha p^\beta=-\mu^2$. It is connected to the nonrelativistic distribution function $f(\vx,\vv)$ in the six-dimensional phase space $(\vx,\vv)$ by \cite{MedeirosDaRosa:2019vkr}
\begin{equation}\label{dist4}
f^{(4)}(x,p)=\mu'^{-3}f(\vx,\vv)\delta(\mu'-\mu)=\mu'^{-3}f(\En,L)\delta(\mu'-\mu),
\end{equation}
%where $f(\En,L)$ is the distribution function in $(x^\alpha,u^\alpha)$ space, and normalized by integrated on full space and hypersurface $g_{\alpha\beta}u^\alpha u^\beta=-1$, the factor $\mu^{-3}$ is from the map between hypersurface $g_{\alpha\beta}p^\alpha p^\beta=-\mu^2$ and $g_{\alpha\beta}u^\alpha u^\beta=-1$.\par
In the second step, Jeans theorem has been used.

Substituting Eqs. \eqref{mvol} and \eqref{dist4} into Eq. \eqref{mcur} and integrating over $\mu'$, we find
\begin{equation}
J^\alpha(x)=2\int\d\En\d L^2\d L_z\ \frac{u^\alpha f(\En,L)}{R(r)^4|u_r||u^\theta|\sin\theta}\sqrt{\frac{B(r)}{A(r)}}.
\end{equation}
Because $(p^t,\pm p^r,\pm p^\theta,p^\phi)$ are mapped to the same $(\En,L^2,L_z,\mu)$, the integral ought to be invariant under reversing the sign of $u^r$ or $u^\theta$. This indicates that $J^r=0$ and $J^\theta=0$, which are in agreement with our assumption that the dark matter is in a steady state. The results are also in agreement with the fact that $J^r$ and $J^\theta$ defined by \eqref{mcur} are integrals of odd functions in symmetric intervals. Similarly, $J^\phi$ vanishes as an integral of an odd function of $L_z$ in a symmetric interval. This is consistent with the assumed spherical symmetry of the distribution of dark matter. The remaining component of the mass-current density four-vector turns out to be
%\begin{eqnarray*}
%J_t&=&-2\int\d\En\d L^2\d L_z\ \frac{\En f(\En,L)}{R(r)^4|u_r||u^\theta|  \sin\theta}\sqrt{\frac{B(r)}{A(r)}}\\
%&=&-2\int\d\En\d L^2\d L_z\ \frac{\En f(\En,L)}{R(r)^2V(r)^{1/2}\left(L^2\sin^2\theta-L_z^2\right)^{1/2}}\\
%&=&-2\pi\int\d\En\d L^2\ \frac{\En f(\En,L)}{R(r)^2V(r)^{1/2}}\\
%&=&-\frac{4\pi}{R(r)^2}\int\d\En\d L\ \frac{\En Lf(\En,L)}{\sqrt{\En^2-A(r)\left[1+L^2/R(r)^2\right]}},
%\end{eqnarray*}
%\begin{eqnarray}\label{flux}
%\nonumber J_t&=&-2\int\d\En\d L^2\d L_z\ \frac{\En f(\En,L)}{R(r)^2V(r)^{1/2}\left(L^2\sin^2\theta-L_z^2\right)^{1/2}}\\
%&=&-\frac{4\pi}{R(r)^2}\int\d\En\d L\ \frac{\En Lf(\En,L)}{V(r)^{1/2}}.
%\end{eqnarray}
\begin{eqnarray}\label{flux}
\nonumber J_t&=&-2\int\d\En\d L^2\d L_z\ \frac{\En f(\En,L)}{R(r)^2V(r)^{1/2}\left(L^2\sin^2\theta-L_z^2\right)^{1/2}}\\
&=&-\frac{4\pi}{R(r)^2}\int\d\En\d L\ \frac{\En Lf(\En,L)}{V(r)^{1/2}}.
\end{eqnarray}

\subsection{Adiabatic invariants}\label{subsect-adinvar}
As is well known in classical mechanics, action variables are adiabatic invariants, i.e., well-preserved approximate constants of a slowly evolving dynamical system. A nonrelativistic analysis in Ref. \cite{Young:1994ed} revealed that the adiabatic invariance of action variables leads to the invariance of the distribution function for an adiabatic growth of a black hole inside a cluster. This was generalized to a relativistic analysis in Ref. \cite{Sadeghian:2013bga} for the Schwarzschild black hole, and in Ref. \cite{Ferrer:2017xwm} for the Kerr black hole.

In this subsection, we will prove that the distribution function of dark matter is adiabatically invariant under the adiabatic growth of a black hole of the general form \eqref{metric-sph}. Based on this result, it is reasonable to match the distribution functions of the initial and the final states if the black hole grows adiabatically in the center.

In a spherical spacetime described by Eq. \eqref{metric-sph} whose parameters are changing adiabatically, there are three action variables for a dark matter particle of mass $\mu$ moving in a bound orbit. They are
\begin{eqnarray}
\label{Ir}I_r(\En,L)\equiv\oint u_r\d r&=&\oint\d r\sqrt{\frac{B(r)}{A(r)}}V(r)^{1/2},\\
\nonumber I_\theta(L,L_z)\equiv\oint u_\theta\d\theta&=&\oint\d\theta(L^2-L_z^2\sin^{-2}\theta)^{1/2}\\
&=&2\pi(L-|L_z|),\\
I_\phi(L_z)\equiv\oint u_\phi\d\phi&=&2\pi L_z,
\end{eqnarray}
where we have neglected an overall factor of $\mu/(2\pi)$. Initially, in the absence of black hole, Eq. \eqref{metric-sph} describes the gravitational field of the preexisting dark matter with a nonrelativistic distribution function $f'(\En',L')$, and the action variables take the form
\begin{eqnarray}
\label{Ipr}\nonumber I'_r(\En',L')&=&\oint\d r\left\{\frac{B'(r)}{A'(r)}\En'^2-B'(r)\left[1+\frac{L'^2}{R'(r)^2}\right]\right\}^{1/2}\\
&\simeq&\oint\d r\left(2E-2\Phi-\frac{L'^2}{r^2}\right)^{1/2},\\
I'_\theta(L',L'_z)&=&2\pi(L'-|L'_z|),\quad I'_\phi(L'_z)=2\pi L'_z.
\end{eqnarray}
In Eq. \eqref{Ipr} we have taken $R'(r)=r$ and the nonrelativistic limit $E=\En'-1\ll1$, $2\Phi(r)=A'(r)-1=B'(r)^{-1}-1\ll1$. Here $E$ is the nonrelativistic energy per unit mass, and $\Phi(r)$ is the Newtonian gravitational potential. Adiabatic invariance of the action variables then suggest that $L=L'$, $L_z=L'_z$ and
\begin{equation}
\frac{\partial I_r(\En,L)}{\partial\En}=\frac{\partial I'_r(\En',L)}{\partial\En'}\frac{\partial\En'}{\partial\En},
\end{equation}
in which the partial derivative of Eq. \eqref{Ir} is
%\begin{equation}
%\frac{\partial I_r(\En,L)}{\partial\En}=\oint\d r\ \En\sqrt{\frac{B(r)}{A(r)}}V(r)^{-1/2}
%\end{equation}
\begin{eqnarray}\label{dIr}
\nonumber\frac{\partial I_r(\En,L)}{\partial\En}&=&\oint\d r\ \frac{\En\sqrt{B(r)/A(r)}}{\sqrt{\En^2-A(r)\left[1+L^2/R(r)^2\right]}}\\
&=&2\int\d r\ \frac{\En\sqrt{B(r)/A(r)}}{\sqrt{\En^2-A(r)\left[1+L^2/R(r)^2\right]}}
\end{eqnarray}
and $\partial I'_r(\En',L)/\partial\En'$ has a similar form.

Consider a hypersurface of constant time, $\Sigma$, to which the normal vector is $\bar{u}=-\left(-g^{tt}\right)^{-1/2}\d t$. The mass enclosed in this hypersurface is \cite{Ferrer:2017xwm,Debbasch:2009uac}
\begin{eqnarray}\label{M}
\nonumber M&=&-\int_\Sigma\d S_\alpha J^\alpha\\
\nonumber&=&\int_\Sigma\d^3x\sqrt{-g}\delta_\alpha^0J^\alpha\\
\nonumber&=&-\int_\Sigma\d r\d\theta\d\phi\sqrt{A(r)B(r)}R(r)^2\sin\theta\frac{J_t}{A(r)}\\
\nonumber&=&\int_\Sigma\d r\int\d\En\d L\ \frac{16\pi^2\En Lf(\En,L)\sqrt{B(r)/A(r)}}{\sqrt{\En^2-A(r)\left[1+L^2/R(r)^2\right]}}\\
&=&\int\d\En\d L\ 8\pi^2Lf(\En,L)\frac{\partial I_r(\En,L)}{\partial\En}.
\end{eqnarray}
In the last line, we have made use of Eq. \eqref{dIr} and neglected the contribution from dark matter particles in unbound orbits. For the initial state, the mass enclosed in the same hypersurface is
\begin{eqnarray}\label{Mp}
\nonumber M&=&\int\d\En'\d L\ 8\pi^2Lf'(\En',L)\frac{\partial I'_r(\En',L)}{\partial\En'}\\
\nonumber&=&\int\d\En\d L\ 8\pi^2Lf'(\En',L)\frac{\partial I'_r(\En',L)}{\partial\En'}\frac{\partial\En'}{\partial\En}\\
&=&\int\d\En\d L\ 8\pi^2Lf'(\En',L)\frac{\partial I_r(\En,L)}{\partial\En}.
\end{eqnarray}
Note here the Jacobian
\begin{equation}
\frac{\partial(\En',L)}{\partial(\En,L)}=
\left|\begin{array}{cc}
\frac{\partial\En'}{\partial\En}\quad&\frac{\partial\En'}{\partial L}\\[5pt]
0\quad&1
\end{array}\right|
=\frac{\partial\En'}{\partial\En}.
\end{equation}
From Eqs. \eqref{M} and \eqref{Mp}, one can conclude directly $f(\En,L)=f'(\En',L)$. That is to say, the distribution function of dark matter is an adiabatic invariant as expected. This paves the way to deriving the final distribution of dark matter $f(\En,L)$ from an initial distribution $f'(\En',L)$, where the argument $\En'$ is a function of $\En$ and $L$ according to the invariance of the radial action variable, $I_r(\En,L)=I'_r(\En',L)$. We should warn that in many situations $\En'(\En,L)$ is a function without an analytic expression, because usually the integrals in Eqs. \eqref{Ir} and \eqref{Ipr} cannot be solved analytically. Numerical methods should then be implemented. Restricted to bound orbits outside the black hole, or roughly $A(r)>0$ and $B(r)>0$, the interval of integration for Eq. \eqref{Ir} is determined by $V(r)\geq0$ with the limits specified by $V(r)=0$, and similarly for Eq. \eqref{Ipr}. This follows from the mathematical requirement that the expression under the square root is nonnegative or the physical requirement that the radial velocity is real. For each bound orbit, the lower and upper limits correspond to the inner and outer turning points.

\subsection{Density profile}\label{subsect-dens}
By definition $J^\alpha=\rho\bar{u}^\alpha$, in which $\bar{u}^\alpha$, the local average four-velocity of dark matter particles in a steady-state distribution, can be regarded as the four-velocity of stationary observers and it is normalized as $\bar{u}_\alpha\bar{u}^\alpha=-1$, so the mass density of dark matter is
\begin{equation}\label{dens}
\rho=-\bar{u}_\alpha J^\alpha=-J_t\sqrt{-g^{tt}}=-J_t/\sqrt{A(r)}.
\end{equation}
For a given black hole geometry and a known dark matter distribution, the density profile can be then calculated in accordance with Eqs. \eqref{flux} and \eqref{dens}, at least in principle.

The integration region for Eq. \eqref{flux} in $\En$-$L$ space can be studied following the same requirements mentioned at the end of the previous subsection. In the integration region, each pair of $(\En,L)$ should allow at least one bound orbit with the radial coordinate $r\in[r_-,r_+]$ determined by $V(r)\geq0$. In other words, the equation $V(r)=0$ should have at least two adjacent roots $r=r_\pm$ outside the horizon and, what is more, the roots should satisfy $\d V(r)/\d r|_{r=r_+}\leq0$ and $\d V(r)/\d r|_{r=r_-}\geq0$. Generally, the existence of bound orbit can be examined by numerical simulations \cite{Ferrer:2017xwm}. In some special cases, analytic constraints can be derived \cite{Sadeghian:2013laa,Will:2012kq}. Fortunately, in applications to the Schwarzschild-like black holes, the integration region in $\En$-$L$ space can be worked out analytically, as we will show in the coming section.

%assign the value of the largest (or second largest) root of $V(r)=0$ to $r_+$ and the value of the second largest (or third largest) root of $V(r)=0$ to $r_-$ if $V(+\infty)<0$ (or $V(+\infty)>0$), and then check the conditions $\d V(r)/\d r|_{r=r_+}\leq0$ and $\d V(r)/\d r|_{r=r_-}\geq0$.

%Solved the problem how to get the final distribution, the question remained is how to know the regions of integration Eq.\eqref{flux}. For a given $r$, an allowable pair of $(\En,L)$ must ensures the corresponding trajectory of particle exists and is bounded (actually, the regions of integration are determined by the form of $V(r)$). All pairs of $(\En,L)$ compose the regions of integration in $\En-L$ space. For an arbitrary spherically symmetric metric, the regions of integration need to be solved accordingly. However, most integration regions could not be solved analytically. For Schwarzschild metric, the integration regions have been solved analytically by L. Sadeghian, F. Ferrer and C. M. Will in Ref. \cite{Sadeghian:2013laa}.

\section{Applications to Schwarzschild-like black holes}\label{sect-Slike}
%\subsection{Schwarzschild black holes in Lorentz-violating backgrounds}\label{subsect-bh}
Lorentz symmetry is of fundamental significance in modern physics. The breaking of Lorentz symmetry has been extensively studied in the literature. Notably, Ref. \cite{Kostelecky:2003fs} pointed out that Riemann-Cartan geometries are incompatible with explicit Lorentz breaking but compatible with spontaneous Lorentz violation. Two classes of spontaneous Lorentz breaking models are of particular interest. In the first class of models, dubbed bumblebee models \cite{Kostelecky:2003fs}, the Lorentz symmetry is spontaneously broken by a vector field (bumblebee field) nonminimally coupled to gravity \cite{Kostelecky:1988zi}. In the second class of models, known as Kalb-Ramond models \cite{Altschul:2009ae}, the spontaneous symmetry breaking arises from a rank-two antisymmetric tensor field (Kalb-Ramond field) nonminimally coupled to gravity. When the bumblebee/Kalb-Ramond field acquires a nonzero vacuum expectation value, the Lorentz symmetry is spontaneously broken in these models.

In the last few years, several spherical static solutions have been obtained from the gravity sector in these models \cite{Casana:2017jkc,Yang:2023wtu,Liu:2024oas}. The solutions resemble the Schwarzschild black hole. In the current section, we will recollect these solutions, prove their equivalence by simple redefinitions of parameters and coordinates, and then apply the framework developed in Sec. \ref{sect-sph} to them.
%In accordance with Refs. \cite{Casana:2017jkc,Yang:2023wtu,Liu:2024oas,Masood:2024oej}, we will assume that the coupling constant between the Ricci tensor and the bumblebee/Kalb-Ramond field is positive.

\begin{enumerate}[(A)]\itshape
\item\label{list-Casana} Bumblebee black hole (BBH)
\end{enumerate}
The line element of a bumblebee black hole is given by \cite{Casana:2017jkc}
\begin{eqnarray}\label{metric-Casana}
\nonumber\d s^2&=&-\left(1-\frac{2Gm}{r}\right)\d t^2+\left(1+\ell\right)\left(1-\frac{2Gm}{r}\right)^{-1}\d r^2\\
&&+r^2\dom
\end{eqnarray}
with the horizon radius $\rh=2Gm$. Here the parameter $\ell=\xi\left|b_\alpha b^\alpha\right|$, in which $b_\alpha$ is the vacuum expectation value of the bumblebee field, and $\xi$ is a real coupling constant between the Ricci tensor and the bumblebee field.
\begin{enumerate}[(B)]\itshape
\item\label{list-Yang} Kalb-Ramond black hole: version 1 (KRBHv1)
\end{enumerate}
The first version of the line element of a Kalb-Ramond black hole is given by \cite{Yang:2023wtu}
\begin{eqnarray}\label{metric-Yang}
\nonumber\d s^2&=&-\left(\frac{1}{1-\ell}-\frac{2Gm}{r}\right)\d t^2+\left(\frac{1}{1-\ell}-\frac{2Gm}{r}\right)^{-1}\d r^2\\
&&+r^2\dom
\end{eqnarray}
with the horizon radius $\rh=2Gm(1-\ell)$. Here $\ell=\xi\left|b_{\alpha\beta}b^{\alpha\beta}\right|/2$, in which $b_{\alpha\beta}$ is the vacuum expectation value of the Kalb-Ramond field, and $\xi$ is a real coupling constant between the Ricci tensor and the Kalb-Ramond field.
\begin{enumerate}[(C)]\itshape
\item\label{list-Liu} Kalb-Ramond black hole: version 2 (KRBHv2)
\end{enumerate}
The second version of the line element of a Kalb-Ramond black hole is given by \cite{Liu:2024oas}
\begin{eqnarray}\label{metric-Liu}
\nonumber\d s^2&=&-\left(1-\frac{2Gm}{r}\right)\d t^2+(1-\ell)\left(1-\frac{2Gm}{r}\right)^{-1}\d r^2\\
&&+r^2\dom
\end{eqnarray}
with the horizon radius $\rh=2Gm$. Here the parameter $\ell$ has the same definition as that in KRBHv1.

Let us take a closer look at these solutions. All of the solutions are Schwarzschild-like but deformed by a parameter $\ell$. If the black hole mass $m$ is set to zero, they will all deviate from Minkowski spacetime unless either the nonminimal coupling constant or the vacuum expectation value of the bumblebee/Kalb-Ramond field is zero. Especially, there will be a geometric singularity in the origin of three-dimensional space similar to a conical singularity in two-dimensional space. For a positive value of deformation parameter $\ell$, BBH is ``stretched'' in the radial direction, while KRBHv1 and KRBHv2 are radially ``contracted''.

Eqs. \eqref{metric-Casana} and \eqref{metric-Liu} are almost the same except for the sign before $\ell$. More interestingly, Eq. \eqref{metric-Liu} can be reproduced from Eq. \eqref{metric-Yang} by replacements $m\rightarrow m/(1-\ell)$, $t\rightarrow t\sqrt{1-\ell}$. Therefore, if we are only interested in gravitational fields, then the three line elements listed above are equivalent. In the rest of this paper, we will restrict our study to KRBHv2 without loss of generality.

Utilizing Eqs. \eqref{flux} and \eqref{Ir}, we obtain for KRBHv2
\begin{eqnarray}
\label{flux-Liu}J_t&=&-\frac{4\pi}{r^2}\int\d\En\d L\ \frac{\En Lf(\En,L)}{V(r)^{1/2}},\\
I_r(\En,L)&=&\sqrt{1-\ell}\oint\d r\ \frac{V(r)^{1/2}}{1-2Gm/r},\\
\label{V-Liu}V(r)&=&\En^2-\left(1-\frac{2Gm}{r}\right)\left(1+\frac{L^2}{r^2}\right).
\end{eqnarray}
%\begin{equation}
%J_t=-\frac{4\pi}{r^2}\int\d\En\d L\ \frac{\En Lf(\En,L)}{V(r)^{1/2}},\quad I_r(\En,L)=\sqrt{1-\ell}\oint\d r\ \frac{V(r)^{1/2}}{1-2Gm/r},\quad V(r)=\En^2-\left(1-\frac{2Gm}{r}\right)\left(1+\frac{L^2}{r^2}\right).
%\end{equation}

As elaborated in Sec. \ref{subsect-dens}, the potential $V(r)$ determines the integration region of Eq. \eqref{flux} in $\En$-$L$ space. Here $V(r)$ has exactly the same form as that for a Schwarzschild black hole, so the integration region of $J_t$ in $\En$-$L$ space is the same as the one for a Schwarzschild black hole \cite{Sadeghian:2013laa}:
\begin{eqnarray}
\label{Lmin}L_\min^2&=&\frac{32(Gm)^2}{36\En^2-27\En^4-8+\En(9\En^2-8)^{3/2}},\\
\label{Lmax}L_\max&=&r\left(\frac{\En^2}{1-2Gm/r}-1\right)^{1/2},\\
\label{Emin}\En_\min&=&\left\{\begin{array}{ll}
\frac{1-2Gm/r}{\sqrt{1-3Gm/r}}\quad&\mathrm{for}~4Gm\leq r\leq6Gm,\\[10pt]
\frac{1+2Gm/r}{\sqrt{1+6Gm/r}}\quad&\mathrm{for}~r\geq6Gm,
\end{array}\right.\\
%\En_\min&=&\left\{
%	\begin{aligned}
%		&(1+2Gm/r)/(1+6Gm/r)^{1/2}:r\geq6Gm\\
%		&(1-2Gm/r)/(1-3Gm/r)^{1/2}:4Gm\leq r\leq 6Gm
%	\end{aligned}
%	\right.\\
\label{Emax}\En_\max&=&1.
\end{eqnarray}
When $r\leq4Gm$, the integration region vanishes. Indeed, as implied by Eq. \eqref{Emin}, $\En_\min=1$ in the limit $r=4Gm$.

In Sec. \ref{subsect-adinvar}, we have verified that the distribution function of dark matter is an adiabatic invariant. As a result, in Eq. \eqref{flux-Liu}, the final distribution function of dark matter around a black hole can be replaced by the initial distribution function in the absence of the black hole. By applying Eq. \eqref{dens}, we find the final dark matter density around KRBHv2 is
%\begin{widetext}
%\end{widetext}
\begin{eqnarray}\label{dens-Liu}
\nonumber\rho(r,\ell)&=&\frac{4\pi}{r^2\sqrt{1-2Gm/r}}\int_{\En_\min(r)}^{1}\En\d\En\int_{L_\min(\En)}^{L_\max(\En,r)}L\\
&&\times\d L\ \frac{f'(E(\En,L,\ell),L)}{\sqrt{\En^2-(1-2Gm/r)(1+L^2/r^2)}},
\end{eqnarray}
where $f'(E,L)$ is the initial distribution function of dark matter. In addition, $E(\En,L,\ell)$ is determined by the adiabatic invariance of the radial action variable,
\begin{eqnarray}\label{E-Liu}
&&\int_{r'_-}^{r'_+}\d r\sqrt{2E-2\Phi(r)-\frac{L^2}{r^2}}\\
\nonumber&=&\sqrt{1-\ell}\int_{r_-}^{r_+}\d r\frac{\sqrt{\En^2-(1-2Gm/r)(1+L^2/r^2)}}{1-2Gm/r},	
\end{eqnarray}
where $r_\pm'$ and $r_\pm$ are respectively the turning points of the initial and final trajectories of the dark matter particle, see our discussion at the end of Sec. \ref{subsect-adinvar}. We have explicitly incorporated $\ell$ into the arguments of $\rho$ and $E$ to highlight their possible dependence on it.

Now let us proceed to compute the dark matter density around KRBHv2 that evolves from two specific initial distributions, i.e., the constant distribution and the Hernquist profile separately.

\subsection{Constant distribution model}\label{subsect-con}
If the initial dark matter distribution is a constant $f'(E,L)=f_0$, then according to the adiabatic invariance of distribution function, the final distribution can be read off directly, $f(\En,L)=f_0$. The final density profile of dark matter turns out to be
\begin{eqnarray}\label{dens-con}
\nonumber\rho(r)&=&\frac{4\pi}{r^2\sqrt{1-2Gm/r}}\int_{\En_\min(r)}^{1}\En\d\En\int_{L_\min(\En)}^{L_\max(\En,r)} L\\
&&\times\d L\ \frac{f_0}{\sqrt{\En^2-(1-2Gm/r)(1+L^2/r^2)}}\\
%\nonumber&=&\frac{4\pi}{r^2\sqrt{1-2Gm/r}}\int_{\En_\min(r)}^{1} \frac{f_0 r^2\sqrt{\En^2-(1-2Gm/r)\left[1+L_\min(\En)^2/r^2\right]}}{1-2Gm/r}\ \En\d\En\\
\nonumber&=&\frac{4\pi f_0}{(1-2Gm/r)^{3/2}}\int_{\En_\min(r)}^{1}\En\\
\nonumber&&\times\d\En\ \sqrt{\En^2-(1-2Gm/r)\left[1+L_\min(\En)^2/r^2\right]}
\end{eqnarray}
in this model, with $L_\min(\En)^2$ and $\En_\min(r)$ given by Eqs. \eqref{Lmin} and \eqref{Emin}.

\begin{figure}[htbp]
	\centering
	\includegraphics[width=0.45\textwidth]{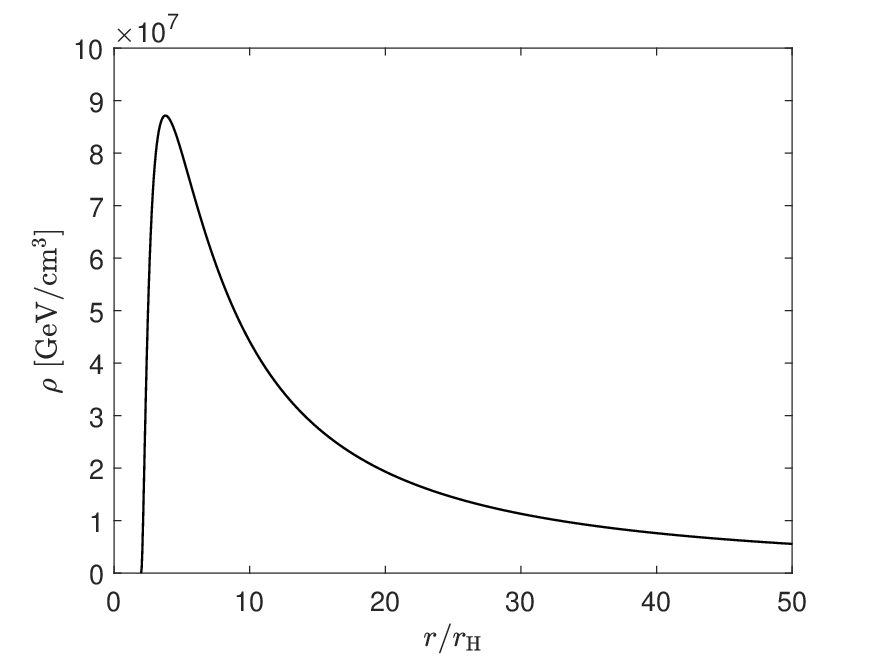}
	\caption{The density profile of dark matter around KRBHv2 evolving from an initial constant distribution. The curve is simulated numerically according to Eq. \eqref{dens-con}.
	}\label{fig:con}
\end{figure}

It is noteworthy that Eqs. \eqref{Lmin}, \eqref{Emin} and \eqref{dens-con} are all independent of the deformation parameter $\ell$ in the line element of KRBHv2. Consequently, the final density profile of dark matter around KRBHv2 ought to be same as the Schwarzschild counterpart, regardless of the value of $\ell$. We plot the profile Eq. \eqref{dens-con} in Fig. \ref{fig:con}, in which we have set $f_0=\rho_0\left(2\pi\sigma_v^2\right)^{-3/2}$ with $\rho_0=0.3\mathrm{GeV}/\mathrm{cm}^3$ and $\sigma_v=100\mathrm{km}/\mathrm{s}$ inferred from the Milky Way \cite{Sadeghian:2013laa}.

\subsection{Hernquist model}\label{subsect-Hern}
The Hernquist model \cite{Hernquist:1990be} is a spherically symmetric matter distribution for which the density and Newtonian gravitational potential are given by
\begin{equation}\label{dens-Hern}
\rho(r)=\frac{\rho_0}{(r/a)(1+r/a)^3},\quad\Phi(r)=-\frac{GM}{a+r},
\end{equation}
where $\rho_0$ and $a$ are characteristic density and radius, respectively, and $M=2\pi\rho_0a^3$ is the total mass of the cluster. This model has the advantage that its associated ergodic distribution function can be found analytically. The distribution function consistent with this potential is \cite{Sadeghian:2013laa}
\begin{eqnarray}\label{dist-Hern}
\nonumber f'(E,L)&=&f_H(E)\\
&=&\frac{M}{\sqrt{2}(2\pi)^3(GMa)^{3/2}}\frac{\sqrt{\tilde{\epsilon}}}{(1-\tilde{\epsilon})^2}\\
\nonumber&&\times\left[(1-2\tilde{\epsilon})\left(8\tilde{\epsilon}^2-8\tilde{\epsilon}-3\right)+\frac{3\arcsin\sqrt{\tilde{\epsilon}}}{\sqrt{\tilde{\epsilon}(1-\tilde{\epsilon})}}\right]
\end{eqnarray}
with $\tilde{\epsilon}=-aE/GM$. In this subsection, we will set $M=10^{12}M_\odot$, $a=20\mathrm{kpc}$, $m=4\times10^6M_\odot$ in our numerical simulations.
%and $f_H(E)$ which proper normalized satisfies $\rho(r)=\int f_H(E)\,\d^3 v$.\par

\begin{figure}[htbp]
	\centering
	\includegraphics[width=0.45\textwidth]{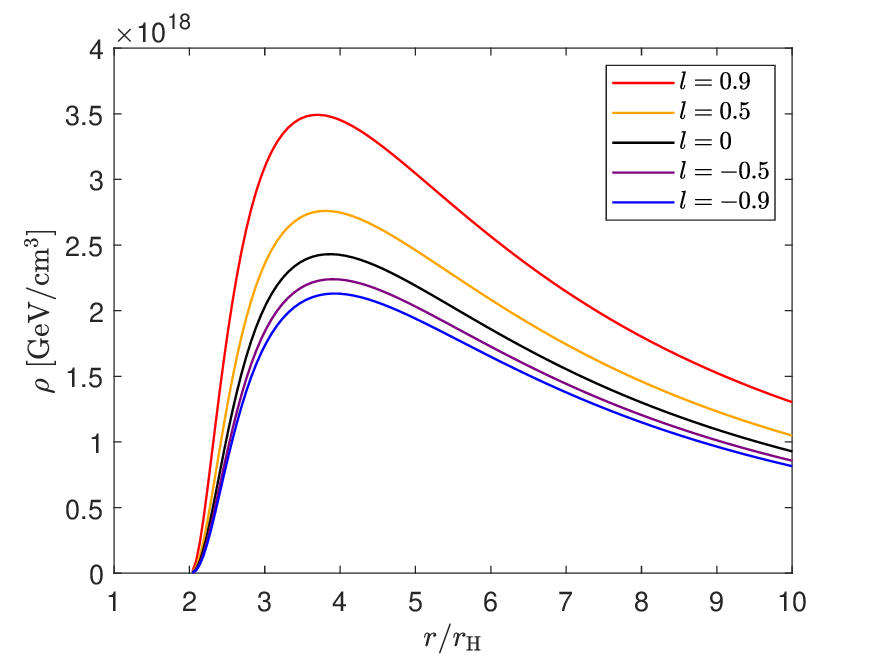}
	\caption{The final density profile of dark matter around KRBHv2 evolving initially from Hernquist distribution.
	}\label{fig:Hern}
\end{figure}
Substituting Eq. \eqref{dist-Hern} into Eq. \eqref{dens-Liu}, and Eq. \eqref{dens-Hern} into Eq. \eqref{E-Liu}, we can follow the numerical procedure and techniques in Ref. \cite{Sadeghian:2013laa} to calculate the final dark matter density around KRBHv2. The major difference is the appearance of parameter $\ell$ in Eq. \eqref{E-Liu}, which eventually introduces dependence of the final density on $\ell$. As illustrated in Fig. \ref{fig:Hern}, both the height and the location of the dark matter spike near KRBHv2 are affected by the value of $\ell$, that is to say, by the deformation of space. If the space is contracted in the radial direction ($\ell>0$), the dark matter spike will be enhanced and dragged closer to the black hole. Oppositely, if the space is stretched in the radial direction ($\ell<0$), the dark matter spike will be depressed and pushed away from the black hole.

In Ref. \cite{Yang:2023wtu}, experimental constraints on $\ell$ have been obtained under the assumption that the gravitational field near the Sun is given by Eq. \eqref{metric-Yang}. Similarly, Ref. \cite{Junior:2024ety} made use of Eq. \eqref{metric-Yang} to describe the gravitational fields near the Earth and Sgr A* and put more experimental constraints on $\ell$. Their results are relegated in Table \ref{tab-ell}. In the planetary, solar and galactic tests, the mass parameter $m$ in Eq. \eqref{metric-Yang} is taken to be the mass of the Earth, the Sun and Sgr A*, respectively. In Ref. \cite{Capozziello:2023tbo}, some of such tests have been implemented to put constraints on $\ell$ by studying Kerr-like black holes surrounded by dark matter spikes in bumblebee gravity.
\begin{table*}[ht]
\caption{Experimental constraints on deformation parameter $\ell$ in KRBHv1 reported by Refs. \cite{Yang:2023wtu,Junior:2024ety}.}\label{tab-ell}
\begin{ruledtabular}
\begin{tabular*}{0.8\textwidth}{llcc}
%\toprule
~ & Tests & Constraints & References \\
\colrule%\midrule
Planetary test & Gyroscope drift & $-6.30714\times10^{-12}\leq\ell\leq3.90708\times10^{-12}$ & \cite{Junior:2024ety}\\
\colrule
\multirow{3}{*}{Solar tests} & Mercury precession & $-3.7\times10^{-12}\leq\ell\leq1.9\times10^{-11}$ & \cite{Yang:2023wtu}\\
& Light deflection & $-1.1\times10^{-10}\leq\ell\leq5.4\times10^{-10}$ & \cite{Yang:2023wtu}\\
& Shapiro time delay & $-6.1\times10^{-13}\leq\ell\leq2.8\times10^{-14}$ & \cite{Yang:2023wtu}\\
\colrule
\multirow{2}{*}{Galactic tests} & S2 star precession & $-0.185022\leq\ell\leq0.060938$ & \cite{Junior:2024ety}\\
& Black hole shadow & $-0.0700225\leq\ell\leq0.189785$ & \cite{Junior:2024ety}\\
%\bottomrule
\end{tabular*}
\end{ruledtabular}
\end{table*}

\begin{figure}[htbp]
    \centering
    \includegraphics[width=0.45\textwidth]{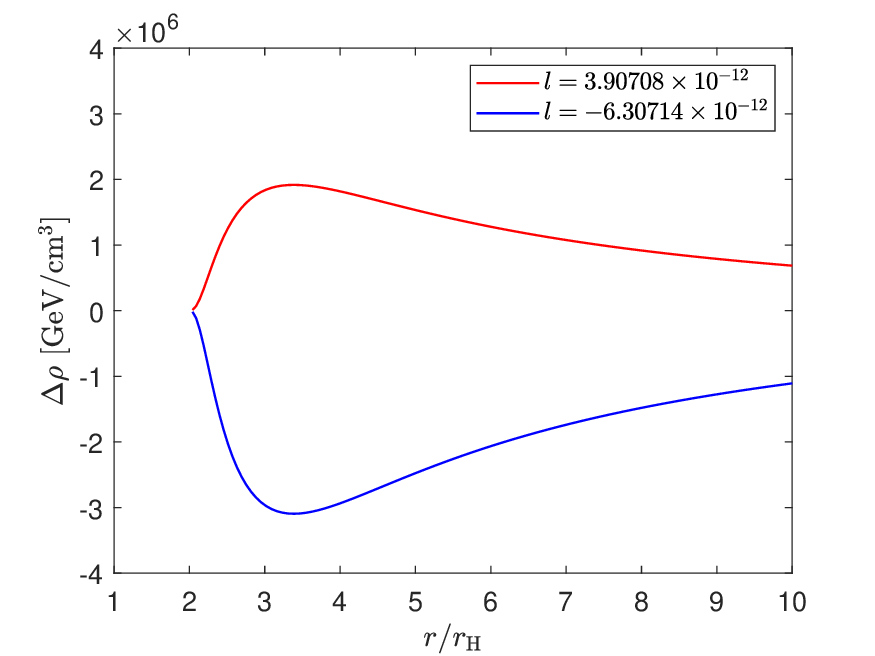}
    \caption{Bounds on the $\ell$-correction to dark matter density based on the planetary test in Table \ref{tab-ell}. The constraints on $\ell$ are drawn from the geodesic drift of gyroscope \cite{Junior:2024ety}.
    }\label{fig:pla}
\end{figure}

\begin{figure}[htbp]
    \centering
    \includegraphics[width=0.45\textwidth]{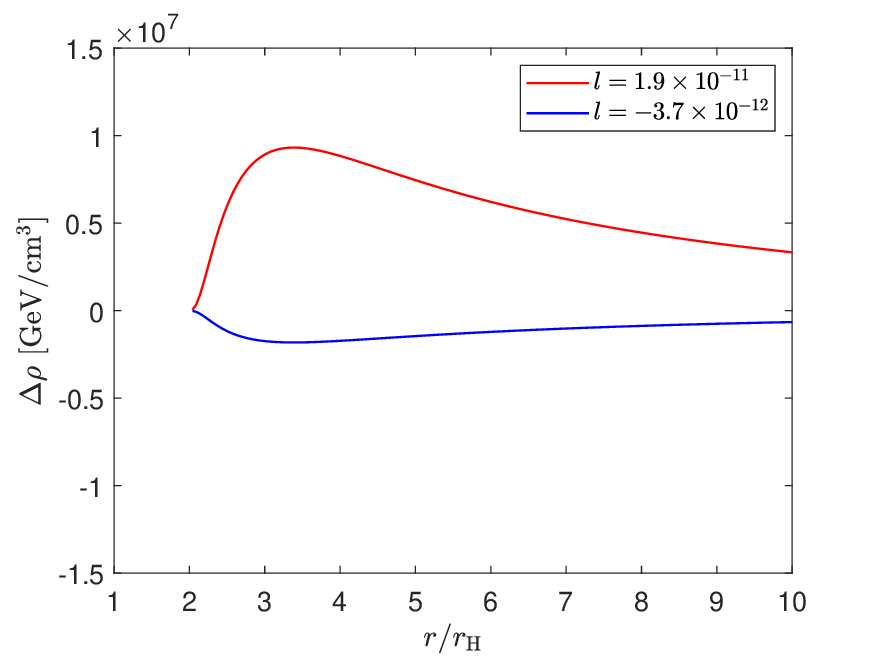}\\
    \includegraphics[width=0.45\textwidth]{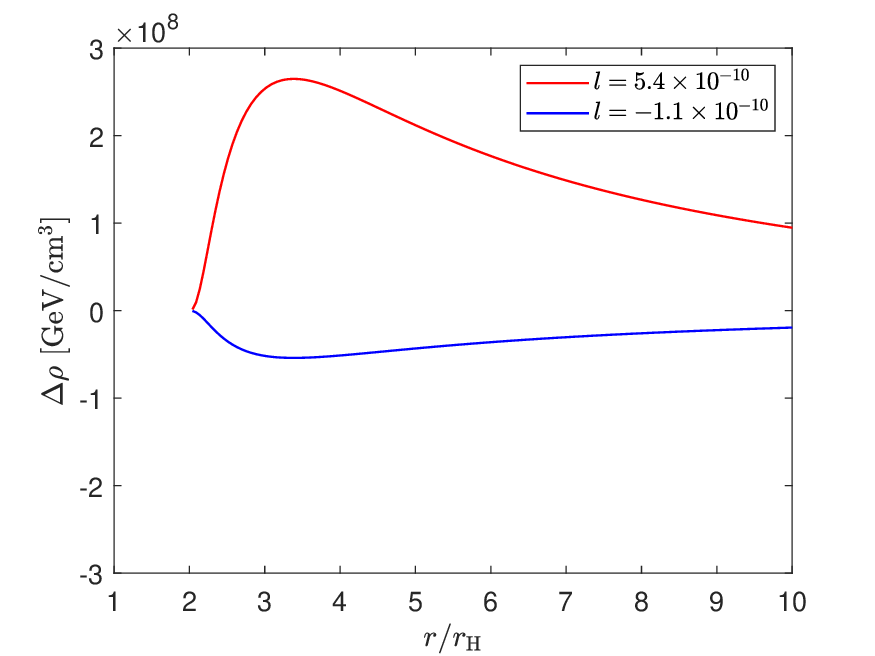}\\
    \includegraphics[width=0.45\textwidth]{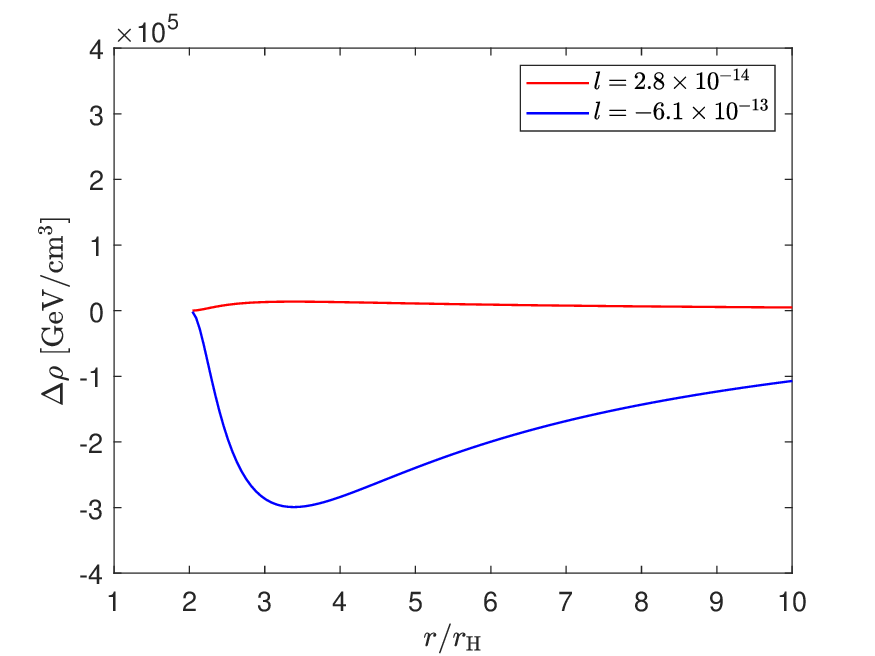}
    \caption{Bounds on the $\ell$-correction to dark matter density based on solar tests in Table \ref{tab-ell}. Constraints on $\ell$ from the Mercury precession, the light deflection and the Shapiro time delay \cite{Yang:2023wtu} are employed in the top, middle and bottom panels, respectively.
    }\label{fig:sol}
\end{figure}

\begin{figure}[htbp]
    \centering
    \includegraphics[width=0.45\textwidth]{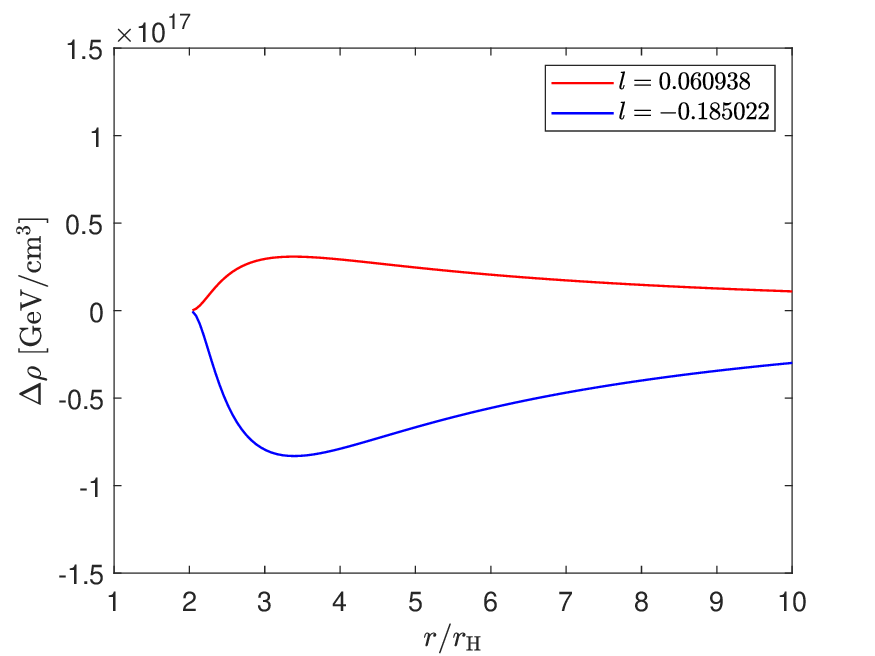}\\
    \includegraphics[width=0.45\textwidth]{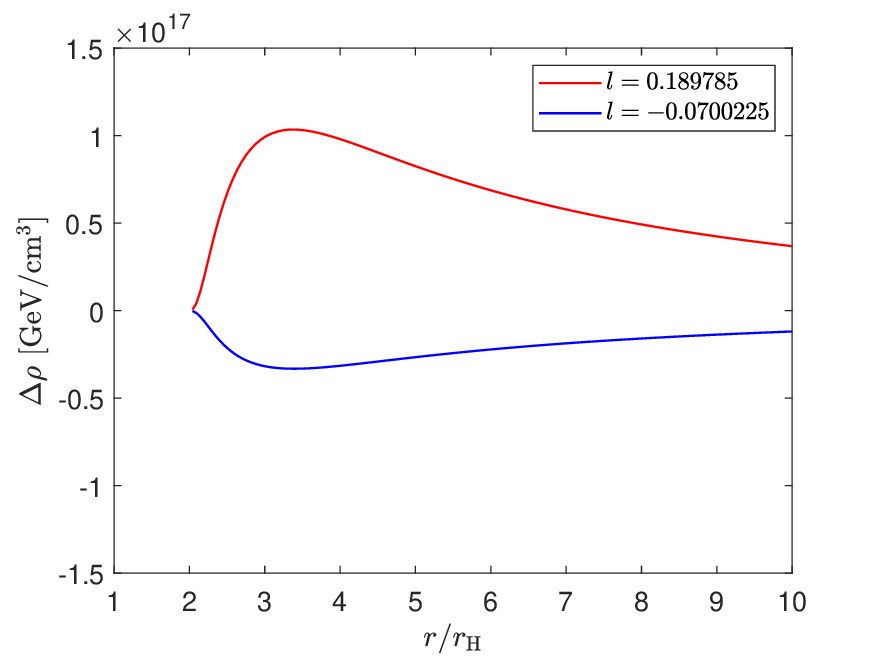}
    \caption{Bounds on the $\ell$-correction to dark matter density based on galactic tests in Table \ref{tab-ell}. The top panel is based on constraints from the S2 star precession, while the bottom panel is based on constraints from the black hole shadow \cite{Junior:2024ety}.
    }\label{fig:gal}
\end{figure}

As we have just mentioned, evolving from Hernquist distribution, the dark matter spike near KRBHv2 is affected by the value of $\ell$. In other words, the deformation parameter $\ell$ induces a correction to the dark matter density profile in this model. Based on experimental constraints in Table \ref{tab-ell}, we simulated bounds on the $\ell$-induced correction to the dark matter spike. The simulation results are shown in Figs. \ref{fig:pla}, \ref{fig:sol} and \ref{fig:gal}, corresponding to planet, solar and galactic tests, respectively. Upper bounds are depicted by red curves, while lower bounds are illustrated by blue curves. According to Fig. \ref{fig:gal}, the $\ell$-induced correction to dark matter density near the spike can reach a few percent. But the bounds are much tighter in Figs. \ref{fig:pla} and \ref{fig:sol}. In these figures, we have assumed that the dark matter density evolves initially from the same Hernquist distribution, and the ordinate is defined as the difference of dark matter densities near KRBHv2 and the Schwarzschild black hole,
\begin{equation}\label{ddens}
\Delta\rho=\rho(r,\ell)-\rho(r,0)=\left.\frac{\partial\rho(r,\ell)}{\partial\ell}\right|_{\ell=0}\times\ell+\mathcal{O}\left(\ell^2\right).
\end{equation}
%In the second step, we have performed the Taylor expansion in powers of $\ell$.
The density difference defined in this way can well characterize the correction induced by $\ell$ to the dark matter spike even if the correction is very small. To improve the numerical efficiency, in Figs. \ref{fig:pla} and \ref{fig:sol} we kept only the first-order term in Eq. \eqref{ddens}, whose coefficient is derived in Appendix \ref{app-nzell}.

\section{Summary}\label{sect-sum}
In the present paper, we extended the relativistic framework of adiabatic growth of Schwarzschild black hole inside dark matter halos \cite{Sadeghian:2013laa} to general static spherical black holes, and we applied it to Schwarzschild-like black holes in bumblebee and Kalb-Ramond models. In detail, we studied whether and how Lorentz-breaking deformations of the Schwarzschild black hole modify the dark matter spike near the black hole.

Surrounding a static spherical black hole with a general line element Eq. \eqref{metric-sph}, we found the density profile of dark matter is related to its distribution function by Eqs. \eqref{V}, \eqref{flux} and \eqref{dens}, and the distribution function is a function of energy and angular momentum. We demonstrated that the distribution function of dark matter is invariant under the adiabatic growth of such a black hole. In addition, the angular momentum is an adiabatic invariant, while the initial and final energies of the dark matter particle are related by the adiabatic invariance of the radial action variable. All together, these results constitute a framework to calculate the density profile of dark matter around a general static spherical black hole from an initial distribution without black hole.

Taking the initial distribution to be either a constant distribution or a Hernquist distribution, we implemented the framework to calculate the density profile of dark matter outside Schwarzschild-like black holes in bumblebee and Kalb-Ramond models. In the literature, there are three solutions of such black holes  \cite{Casana:2017jkc,Yang:2023wtu,Liu:2024oas}, all characterized by a mass parameter and a Lorentz-breaking parameter. We tagged them as BBH, KRBHv1 and KRBHv2 and proved that their line elements are equivalent. Then we simulated the profile of dark matter density around KRBHv2. Similar to the Schwarzschild black hole, KRBHv2 generates a near-horizon spike in the dark matter density profile.

For a constant initial distribution, the final density profile of dark matter outside KRBHv2 is the same as the profile outside the Schwarzschild black hole. However, if the initial distribution of dark matter is the Hernquist distribution, then its final density will depend on the Lorentz-breaking parameter. Although we have restricted our study to KRBHv2, the results apply to all of the three Schwarzschild-like black holes listed in Sec. \ref{sect-Slike}. Making use of experimental constraints on the Lorentz-breaking parameter, we put stringent bounds on difference of dark matter densities near KRBHv2 and Schwarzschild black hole.

As concrete examples, we have chosen to study Schwarzschild-like black holes not because they are more practical, but because they are more tractable numerically. In the future, we will overcome numerical difficulties and apply the framework here to more practical cases such as the Schwarzschild-de Sitter black hole \cite{Masood:2024oej}.

In Sec. \ref{subsect-Hern}, we have made use of the constraints on $\ell$ from the Event Horizon Telescope observations, see Ref. \cite{Junior:2024ety}, which neglected the deformation of the line element by the dark matter spike. As demonstrated in Ref. \cite{Capozziello:2023tbo} for Kerr-like black holes, when taking the dark matter spike into account, the mass parameter $m$ in the line element should be replaced by a spike-dependent function of $r$. In the future, it will be valuable to refine the observational constraints and our results along this line.

\acknowledgments{This work is sponsored by the Natural Science Foundation of Shanghai (Grant No. 24ZR1419300).}

\appendix
\section{Dark matter density at near-zero $\ell$}\label{app-nzell}
In Sec. \ref{subsect-Hern}, we have defined the density difference $\Delta\rho$ between dark matter around KRBHv2 and dark matter around the Schwarzschild black hole. When $\left|\ell\right|\ll1$, it can be well estimated by the linear-order term in Eq. \eqref{ddens},
\begin{equation}\label{ddens1}
\Delta\rho\simeq\left.\frac{\partial\rho(r,\ell)}{\partial\ell}\right|_{\ell=0}\times\ell.
\end{equation}
In order to determine the coefficient in this expression, we try to calculate the partial derivative of Eq. \eqref{dens-Liu} with respect to $\ell$,
\begin{comment}
{\color{red}
\begin{equation}
	\rho(r)=\frac{4\pi}{r^2\sqrt{1-2Gm/r}}\int_{\En_\min(r)}^{1}\En\d\En\int_{L_\min(\En)}^{L_\max(\En,r)}L\d L\ \frac{f^{\prime}(E(\En,L),L)}{\sqrt{\En^2-(1-2Gm/r)(1+L^2/r^2)}},
\end{equation}
\begin{equation}
	\int_{r'_-}^{r'_+}\d r\sqrt{2E-2\Phi(r)-\frac{L^2}{r^2}}=\sqrt{1-\ell}\int_{r_-}^{r_+}\d r\frac{\sqrt{\En^2-(1-2Gm/r)(1+L^2/r^2)}}{1-2Gm/r},	
\end{equation}
\begin{equation}
	\rho(r,\ell)=\frac{4\pi}{r^2\sqrt{1-2Gm/r}}\int_{\En_\min(r)}^{1}\En\d\En\int_{L_\min(\En)}^{L_\max(\En,r)}L\d L\ \frac{f^{\prime}(E(\En,L,\ell),L)}{\sqrt{\En^2-(1-2Gm/r)(1+L^2/r^2)}},
\end{equation}
\begin{equation}
	\frac{\partial E}{\partial\ell}=\frac{\partial I/\partial\ell}{\partial I^{\prime}/\partial E},
\end{equation}
\begin{equation}
	\frac{\partial I}{\partial \ell}=-\frac{1}{\sqrt{1-\ell}}\int_{r_-}^{r_+}\d r\frac{\sqrt{\En^2-(1-2Gm/r)(1+L^2/r^2)}}{1-2Gm/r},
\end{equation}
\begin{eqnarray}
	\frac{\partial I^{\prime}}{\partial E}&=&\frac{\partial}{\partial E}\left( 2\int_{r'_-(E,L)}^{r'_+(E,L)}\d r\sqrt{2E-2\Phi(r)-\frac{L^2}{r^2}}\ \right),\\
	&=&2\sqrt{2E-2\Phi(r'_+)-\frac{L^2}{(r'_+)^2}}\ \frac{\partial r'_+}{\partial E}-2\sqrt{2E-2\Phi(r'_-)-\frac{L^2}{(r'_-)^2}}\ \frac{\partial r'_-}{\partial E}+2\int_{r'_-(E,L)}^{r'_+(E,L)}\d r\frac{1}{\sqrt{2E-2\Phi(r)-\frac{L^2}{r^2}}},\\
	&=&2\int_{r'_-(E,L)}^{r'_+(E,L)}\d r\frac{1}{\sqrt{2E-2\Phi(r)-\frac{L^2}{r^2}}},
\end{eqnarray}
}
\end{comment}
\begin{eqnarray}\label{diffrho}
\nonumber\frac{\partial\rho}{\partial\ell}&=&\frac{4\pi}{r^2\sqrt{1-2Gm/r}}\int_{\En_\min(r)}^{1}\En\d\En\int_{L_\min(\En)}^{L_\max(\En,r)}L\\
&&\times\d L\ \frac{(\partial f'/\partial E)(\partial E/\partial\ell)}{\sqrt{\En^2-(1-2Gm/r)(1+L^2/r^2)}},
\end{eqnarray}
which involves $\partial f'/\partial E$ and $\partial E/\partial\ell$. The latter can be got from differentiation of Eq. \eqref{E-Liu} with respect to $\ell$. Specifically, the left hand side gives
\begin{widetext}
\begin{eqnarray}\label{dleft}
\nonumber&&\frac{\partial E}{\partial\ell}\frac{\partial}{\partial E}\left(\int_{r'_-(E,L)}^{r'_+(E,L)}\d r\sqrt{2E-2\Phi(r)-\frac{L^2}{r^2}}\ \right)\\
\nonumber&=&\left[\sqrt{2E-2\Phi(r'_+)-\frac{L^2}{(r'_+)^2}}\ \frac{\partial r'_+}{\partial E}-\sqrt{2E-2\Phi(r'_-)-\frac{L^2}{(r'_-)^2}}\ \frac{\partial r'_-}{\partial E}+\int_{r'_-(E,L)}^{r'_+(E,L)}\d r\frac{1}{\sqrt{2E-2\Phi(r)-\frac{L^2}{r^2}}}\right]\frac{\partial E}{\partial\ell}\\
&=&\frac{\partial E}{\partial\ell}\int_{r'_-(E,L)}^{r'_+(E,L)}\d r\frac{1}{\sqrt{2E-2\Phi(r)-\frac{L^2}{r^2}}},
\end{eqnarray}
\end{widetext}
while the right hand side yields
\begin{equation}\label{dright}
-\frac{1}{2\sqrt{1-\ell}}\int_{r_-}^{r_+}\d r\frac{\sqrt{\En^2-(1-2Gm/r)(1+L^2/r^2)}}{1-2Gm/r}.
\end{equation}
Equating Eq. \eqref{dleft} to Eq. \eqref{dright}, we can obtain
\begin{eqnarray}
\nonumber\frac{\partial E}{\partial\ell}&=&-\frac{1}{2\sqrt{1-\ell}}\left[\int_{r'_-}^{r'_+}\d r\frac{1}{\sqrt{2E-2\Phi(r)-\frac{L^2}{r^2}}}\right]^{-1}\\
&&\times\int_{r_-}^{r_+}\d r\frac{\sqrt{\En^2-(1-2Gm/r)(1+L^2/r^2)}}{1-2Gm/r}.
\end{eqnarray}
For the constant distribution model, we have $\partial f'/\partial E=0$. In contrast, for the Hernquist model, we find
\begin{eqnarray}
\frac{\partial f^{\prime}}{\partial E}&=&-\frac{Ma^2}{2\sqrt{2}(2\pi)^3(GMa)^{5/2}}\frac{\sqrt{\tilde{\epsilon}}}{(1-\tilde{\epsilon})^3}\\
\nonumber&&\times\left[\left(48\tilde{\epsilon}^3-136\tilde{\epsilon}^2+118\tilde{\epsilon}-15\right)+\frac{15\arcsin(\sqrt{\tilde{\epsilon}})}{\sqrt{\tilde{\epsilon}(1-\tilde{\epsilon})}}\right]
\end{eqnarray}
with $\tilde{\epsilon}=-aE/GM$, and $E$ is dictated by Eq. \eqref{E-Liu}. Evaluating $\partial f'/\partial E$ and $\partial E/\partial\ell$ at $\ell=0$, and inserting them into Eq. \eqref{diffrho}, one can numerically compute $\partial\rho/\partial\ell$ at $\ell=0$, which is nothing else but the coefficient in Eq. \eqref{ddens1}.


\begin{thebibliography}{99}
\bibitem{Bertone:2004pz}
G.~Bertone, D.~Hooper and J.~Silk,
%``Particle dark matter: Evidence, candidates and constraints,''
Phys. Rept. \textbf{405}, 279-390 (2005)
%doi:10.1016/j.physrep.2004.08.031
[arXiv:hep-ph/0404175 [hep-ph]].

\bibitem{Hernquist:1990be}
L.~Hernquist,
%``An Analytical Model for Spherical Galaxies and Bulges,''
Astrophys. J. \textbf{356}, 359 (1990)
%doi:10.1086/168845

\bibitem{Zhao:1995cp}
H.~Zhao,
%``Analytical models for galactic nuclei,''
Mon. Not. Roy. Astron. Soc. \textbf{278}, 488-496 (1996)
%doi:10.1093/mnras/278.2.488
[arXiv:astro-ph/9509122 [astro-ph]].

\bibitem{Gondolo:1999ef}
P.~Gondolo and J.~Silk,
%``Dark matter annihilation at the galactic center,''
Phys. Rev. Lett. \textbf{83}, 1719-1722 (1999)
%doi:10.1103/PhysRevLett.83.1719
[arXiv:astro-ph/9906391 [astro-ph]].

\bibitem{Sadeghian:2013laa}
L.~Sadeghian, F.~Ferrer and C.~M.~Will,
%``Dark matter distributions around massive black holes: A general relativistic analysis,''
Phys. Rev. D \textbf{88}, no.6, 063522 (2013)
%doi:10.1103/PhysRevD.88.063522
[arXiv:1305.2619 [astro-ph.GA]].

\bibitem{Ferrer:2017xwm}
F.~Ferrer, A.~M.~da Rosa and C.~M.~Will,
%``Dark matter spikes in the vicinity of Kerr black holes,''
Phys. Rev. D \textbf{96}, no.8, 083014 (2017)
%doi:10.1103/PhysRevD.96.083014
[arXiv:1707.06302 [astro-ph.CO]].

\bibitem{Ullio:2001fb}
P.~Ullio, H.~Zhao and M.~Kamionkowski,
%``A Dark matter spike at the galactic center?,''
Phys. Rev. D \textbf{64}, 043504 (2001)
%doi:10.1103/PhysRevD.64.043504
[arXiv:astro-ph/0101481 [astro-ph]].

\bibitem{Merritt:2003qk}
D.~Merritt,
%``Evolution of the dark matter distribution at the galactic center,''
Phys. Rev. Lett. \textbf{92}, 201304 (2004)
%doi:10.1103/PhysRevLett.92.201304
[arXiv:astro-ph/0311594 [astro-ph]].

\bibitem{Bertone:2024wbn}
G.~Bertone, A.~R.~A.~C.~Wierda, D.~Gaggero, B.~J.~Kavanagh, M.~Volonteri and N.~Yoshida,
%``Dark Matter Mounds: towards a realistic description of dark matter overdensities around black holes,''
[arXiv:2404.08731 [astro-ph.CO]].

\bibitem{Capozziello:2023rfv}
S.~Capozziello, S.~Zare, D.~F.~Mota and H.~Hassanabadi,
%``Dark matter spike around Bumblebee black holes,''
JCAP \textbf{05}, 027 (2023)
%doi:10.1088/1475-7516/2023/05/027
[arXiv:2303.13554 [gr-qc]].

\bibitem{Shen:2023erj}
Z.~Shen, A.~Wang, Y.~Gong and S.~Yin,
%``Analytical models of supermassive black holes in galaxies surrounded by dark matter halos,''
Phys. Lett. B \textbf{855}, 138797 (2024)
%doi:10.1016/j.physletb.2024.138797
[arXiv:2311.12259 [gr-qc]].

\bibitem{Zhang:2025mdl}
Z.~C.~Zhang, H.~C.~Yuan and Y.~Tang,
%``Universal Density and Velocity Distributions of Dark Matter around Massive Black Holes,''
[arXiv:2503.02573 [astro-ph.GA]].

\bibitem{Casana:2017jkc}
R.~Casana, A.~Cavalcante, F.~P.~Poulis and E.~B.~Santos,
%``Exact Schwarzschild-like solution in a bumblebee gravity model,''
Phys. Rev. D \textbf{97}, no.10, 104001 (2018)
%doi:10.1103/PhysRevD.97.104001
[arXiv:1711.02273 [gr-qc]].

\bibitem{Yang:2023wtu}
K.~Yang, Y.~Z.~Chen, Z.~Q.~Duan and J.~Y.~Zhao,
%``Static and spherically symmetric black holes in gravity with a background Kalb-Ramond field,''
Phys. Rev. D \textbf{108}, no.12, 124004 (2023)
%doi:10.1103/PhysRevD.108.124004
[arXiv:2308.06613 [gr-qc]].

\bibitem{Liu:2024oas}
W.~Liu, D.~Wu and J.~Wang,
%``Static neutral black holes in Kalb-Ramond gravity,''
JCAP \textbf{09}, 017 (2024)
%doi:10.1088/1475-7516/2024/09/017
[arXiv:2406.13461 [hep-th]].

\bibitem{MedeirosDaRosa:2019vkr}
A.~Medeiros Da Rosa,
``Adiabatic Dark Matter Density Cusps Around Supermassive Black Holes and Dark Matter Detection,''
Ph.D. thesis, Washington University in St. Louis, 2019.
%doi:10.7936/205a-j912

\bibitem{Young:1994ed}
P.~Young,
%``Numerical models of star clusters with a central black hole. I - Adiabatic models.,''
Astrophys. J. \textbf{242}, 1232-1237 (1980)
%doi:10.1086/158553

\bibitem{Sadeghian:2013bga}
L.~Sadeghian,
``Star Clusters and Dark Matter as Probes of the Spacetime Geometry of Massive Black Holes,''
Ph.D. thesis, Washington University in St. Louis, 2013.
%doi:10.7936/K7JM27QG
[arXiv:1308.5378 [gr-qc]].

\bibitem{Debbasch:2009uac}
F.~Debbasch and W.~A.~Van Leeuwen,
%``General relativistic Boltzmann equation, I: Covariant treatment,''
Physica A \textbf{388}, 1079-1104 (2009)
%doi:10.1016/j.physa.2008.12.023

\bibitem{Will:2012kq}
C.~M.~Will,
%``Capture of non-relativistic particles in eccentric orbits by a Kerr black hole,''
Class. Quant. Grav. \textbf{29}, 217001 (2012)
%doi:10.1088/0264-9381/29/21/217001
[arXiv:1208.3931 [astro-ph.GA]].

\bibitem{Kostelecky:2003fs}
V.~A.~Kostelecky,
%``Gravity, Lorentz violation, and the standard model,''
Phys. Rev. D \textbf{69}, 105009 (2004)
%doi:10.1103/PhysRevD.69.105009
[arXiv:hep-th/0312310 [hep-th]].

\bibitem{Kostelecky:1988zi}
V.~A.~Kostelecky and S.~Samuel,
%``Spontaneous Breaking of Lorentz Symmetry in String Theory,''
Phys. Rev. D \textbf{39}, 683 (1989)
%doi:10.1103/PhysRevD.39.683

\bibitem{Altschul:2009ae}
B.~Altschul, Q.~G.~Bailey and V.~A.~Kostelecky,
%``Lorentz violation with an antisymmetric tensor,''
Phys. Rev. D \textbf{81}, 065028 (2010)
%doi:10.1103/PhysRevD.81.065028
[arXiv:0912.4852 [gr-qc]].

\bibitem{Junior:2024ety}
E.~L.~B.~Junior, J.~T.~S.~S.~Junior, F.~S.~N.~Lobo, M.~E.~Rodrigues, D.~Rubiera-Garcia, L.~F.~D.~da Silva and H.~A.~Vieira,
%``Spontaneous Lorentz symmetry-breaking constraints in Kalb\textendash{}Ramond gravity,''
Eur. Phys. J. C \textbf{84}, no.12, 1257 (2024)
%doi:10.1140/epjc/s10052-024-13619-3
[arXiv:2405.03291 [gr-qc]].

\bibitem{Capozziello:2023tbo}
S.~Capozziello, S.~Zare, L.~M.~Nieto and H.~Hassanabadi,
%``Modified Kerr black holes surrounded by dark matter spike,''
[arXiv:2311.12896 [gr-qc]].

\bibitem{Masood:2024oej}
S.~Masood and S.~Mikki,
%``The thermodynamic profile of AdS black holes in Lorentz invariance-violating Bumblebee and Kalb-Ramond gravity,''
[arXiv:2411.06188 [gr-qc]].

%\bibitem{Anninos:2008fx}
%D.~Anninos, W.~Li, M.~Padi, W.~Song and A.~Strominger,
%``Warped AdS(3) Black Holes,''
%JHEP \textbf{03}, 130 (2009)
%doi:10.1088/1126-6708/2009/03/130
%[arXiv:0807.3040 [hep-th]].

%\bibitem{Chen:2019nhv}
%B.~Chen, F.~L.~Lin and B.~Ning,
%``Gedanken Experiments to Destroy a BTZ Black Hole,''
%Phys. Rev. D \textbf{100}, no.4, 044043 (2019)
%doi:10.1103/PhysRevD.100.044043
%[arXiv:1902.00949 [gr-qc]].
\end{thebibliography}
\end{document}